\documentclass[aps,prx,reprint]{revtex4-2}

\usepackage{hyperref}
\usepackage{amsmath}
\usepackage{amssymb}
\usepackage{dsfont}
\usepackage[T1]{fontenc}
\usepackage{ae}
\usepackage{color}
\usepackage{graphicx} 
\usepackage[mediumspace,mediumqspace,squaren]{SIunits}

\newcommand{\ee} {{\textrm e}}
\newcommand{\dd} {\hbox{\textrm d}}

\newcommand{\bra}[1]{\ensuremath{\left\langle #1 \right|}}
\newcommand{\ket}[1]{\ensuremath{\left| #1 \right\rangle}}

\newcommand{\oa} {{\hat{a}}}
\newcommand{\ob} {{\hat{b}}}

\newcommand{\os} {{\hat{\sigma}}}
\newcommand{\osd} {{\hat{\sigma}^\dag}}
\newcommand{\oA} {{\hat{A}}}
\newcommand{\oAd} {{\hat{A}^\dag}}
\newcommand{\orho} {{\hat{\rho}}}
\newcommand{\oX} {{\hat{X}}}
\newcommand{\oP} {{\hat{P}}}

\begin{document}

\title{Deterministic Free-Propagating Photonic Qubits with Negative Wigner Functions}

\author{Valentin Magro} \affiliation{JEIP,  UAR 3573 CNRS, Coll{\`e}ge de France, PSL University, 11, place Marcelin Berthelot, 75231 Paris Cedex 05, France}

\author{Julien Vaneecloo} \affiliation{JEIP,  UAR 3573 CNRS, Coll{\`e}ge de France, PSL University, 11, place Marcelin Berthelot, 75231 Paris Cedex 05, France}

\author{S{\'e}bastien Garcia} \affiliation{JEIP,  UAR 3573 CNRS, Coll{\`e}ge de France, PSL University, 11, place Marcelin Berthelot, 75231 Paris Cedex 05, France}

\author{Alexei Ourjoumtsev} \email{Corresponding author: alexei.ourjoumtsev@college-de-france.fr} \affiliation{JEIP,  UAR 3573 CNRS, Coll{\`e}ge de France, PSL University, 11, place Marcelin Berthelot, 75231 Paris Cedex 05, France}

\begin{abstract}

Engineering quantum states of free-propagating light is of paramount importance for quantum technologies. Coherent states ubiquitous in classical and quantum communications, squeezed states used in quantum sensing \cite{Tse2019,Acernese2019}, and even highly-entangled cluster states studied in the context of quantum computing \cite{Larsen2019,Asavanant2019} can be produced deterministically, but they obey quasi-classical optical field statistics described by Gaussian, positive Wigner functions. Fully harnessing the potential of many quantum engineering protocols requires using non-Gaussian Wigner-negative states \cite{Walschaers2021}, so far produced using intrinsically probabilistic methods. Here we describe the first fully deterministic preparation of non-Gaussian Wigner-negative free-propagating states of light, obtained by mapping the internal state of an intracavity Rydberg superatom onto an optical qubit encoded as a superposition of $0$ and $1$ photons. This approach allows us to reach a $60\%$ photon generation efficiency in a well-controlled spatio-temporal mode, while maintaining a strong photon antibunching. By changing the qubit rotation angle, we observe an evolution from quadrature squeezing to Wigner negativity. Our experiment sets this new technique as a viable method to deterministically generate non-Gaussian photonic resources, lifting several major roadblocks in optical quantum engineering.

\end{abstract}

\maketitle

Outside the microwave domain \cite{Bertet2002,Hofheinz2009}, photonic non-Gaussian states are difficult to produce in an efficient, scalable way. Currently, they are generally heralded via photon detections in parametric fluorescence or Raman scattering \cite{Lvovsky2020}, with low probabilities. A major step towards their deterministic generation was made recently, when optical ``Schr\"odinger cats" were prepared using a single atom in a high-finesse cavity \cite{Hacker2019}, but the cat's parity still relied on the random outcome of a measurement. Single quantum emitters, which form, in principle, the simplest deterministic non-Gaussian light sources, made spectacular progress in recent years \cite{Senellart2017}, and measurements of their scattered field statistics revealed quadrature squeezing \cite{Schulte2015}, but to this day losses and noise made states with negative Wigner functions elusive with this approach.

Here, we describe the first fully deterministic preparation of non-Gaussian Wigner-negative free-propagating optical quantum states. In our setup, a small atomic cloud placed inside a medium-finesse optical cavity and driven to a highly-excited Rydberg state acts as a single two-level collective ``superatom''. We coherently control its internal state, then map it onto a free-propagating light pulse to produce an optical qubit  $\cos(\theta/2)\ket{0}+\sin(\theta/2)\ket{1}$ encoded as a quantum superposition of $0$ and $1$ photons. Its single-photon character is revealed by photon correlation measurements showing strong antibunching, while homodyne tomography gives us access to the full quantum statistics of the prepared light fields. In agreement with theoretical predictions, the reconstructed Wigner functions are quadrature-squeezed for small mixing angles $\theta$, and develop a negative region when $\theta$ increases and the one-photon component becomes dominant. The generated states, emitted in the desired spatio-temporal mode with a high $60\%$ efficiency \cite{Tomm2021}, are fully compatible with high-efficiency quantum memories \cite{VernazGris2018,Wang2019} and could boost the performance of photonic quantum engineering platforms.

\section*{Results}

\subsection*{Experimental protocol}
\begin{figure*}[t]
\centering
\includegraphics[width=170mm]{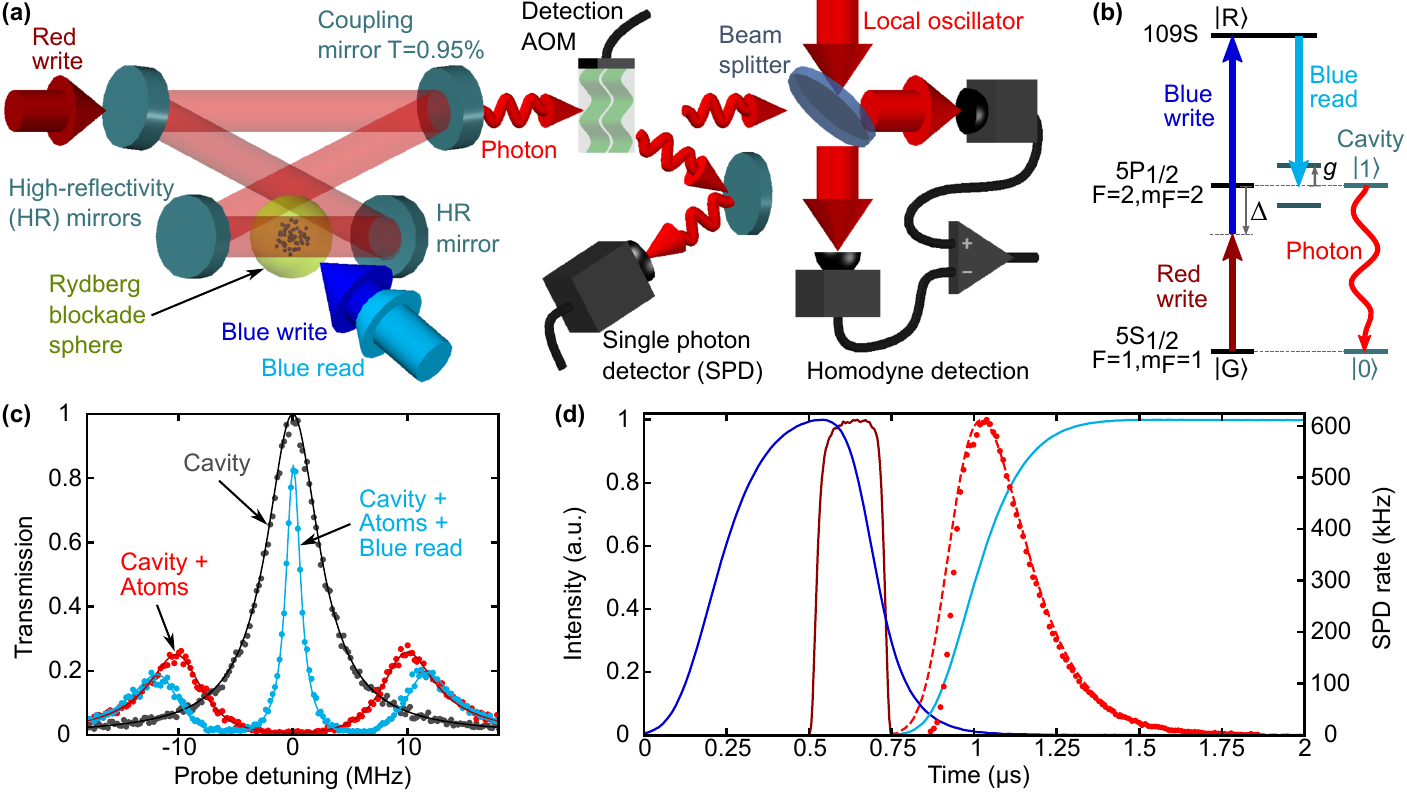}
\caption{ \textbf{Deterministic generation of number-encoded photonic qubits with an intracavity Rydberg polariton.} (a) Schematics of the experimental apparatus featuring a small ensemble of cold atoms inside an optical cavity. The photonic states, released through the cavity coupling mirror, are directed with an acousto-optic modulator (AOM) either towards single photon detectors (SPD) or an homodyne detection. (b) Diagram of energy levels illustrating the off-resonant two photon excitation of a single Rydberg excitation (write step), followed by photonic state release through a cavity-Rydberg polariton (read step). (c) Cavity transmission spectra of a laser probe frequency-detuned around the bare cavity resonance (black). The strong collective coupling of the atomic ensemble to the cavity field yields a vacuum Rabi splitting (red). Additionally coupling to Rydberg level with the blue read beam creates a dark cavity-Rydberg polariton at zero detuning. (d) Measured temporal evolution of normalized laser beam intensities (full lines, color code from (b)) and SPD rate proportional to output photon flux $I(t)$ (red dots, right axis). The dashed line models the photonic state release from the blue read evolution and independently-measured parameters (see text).}
\label{fig:setup}
\end{figure*}

The core of our experimental setup, depicted in Fig.~\ref{fig:setup}(a), features a single Rydberg superatom strongly coupled to a medium-finesse single-ended optical cavity \cite{Vaneecloo2022}. The superatom is made of about $800$ cold $^{87}$Rb atoms initialized in the state \ket{G} (Fig.~\ref{fig:setup}(b)). In the first ``write'' step of our protocol, a \unit{795}{\nano\meter} red write laser and a \unit{475}{\nano\meter} blue write laser drive a resonant two-photon transition to the Rydberg state $R : 109S$, with a detuning $\Delta/(2\pi) = \unit{-500}{\mega\hertz}$ from the intermediate state $E : 5P_{1/2}$.  The \unit{\sim 20}{\micro\meter} Rydberg blockade range~\cite{Tong2004} largely exceeds the $\unit{4.5}{\micro\meter}$ Gaussian rms radius of the cloud, strongly inhibiting multiple Rydberg excitations and turning the cloud into an effective two-level system with a ground state $\ket{G}$ and a collective singly-excited Rydberg state $\ket{R}$. Rabi oscillations between \ket{G} and \ket{R} driven with the write beams allow us to prepare a coherent superposition $\cos(\theta/2)\ket{G}-\sin(\theta/2)\ket{R}$ with a controlled qubit rotation angle $\theta$.

During the following ``read'' step, $\ket{R}$ is mapped via the cavity, resonant with the $G-E$ transition, onto a \unit{795}{\nano\meter} single-mode free-propagating photon with fully controlled spatial, temporal, spectral and polarization degrees of freedom. The strong superatom-cavity coupling strength $g = 2\pi \times \unit{10}{\mega\hertz}$, exceeding the decay rates of both the $G-E$ atomic dipole $\gamma = 2\pi \times \unit{2.87}{\mega\hertz} $ and the cavity field $\kappa = 2\pi \times \unit{2.8}{\mega\hertz}$, results in a vacuum Rabi splitting of two well-separated bright polariton modes detuned by $\pm g$ from the cavity resonance, visible on Fig.~\ref{fig:setup}(c). Adiabatically increasing the Rabi frequency $\Omega$ of the blue read laser beam, coupling $E$ and $R$, opens a resonant electromagnetically-induced transparency window in the spectrum and coherently transfers the excitation from \ket{R} to an intra-cavity photon $\ket{1_{\mathrm{c}}}$ via the rotation of the dark polariton state
\begin{equation}
\ket{D} = \cos\beta \, \ket{G}\ket{1_{\mathrm{c}}} - \sin\beta \, \ket{R}\ket{0_{\mathrm{c}}} \ ,
\end{equation}
with the angle $\beta = \arctan(2 g/\Omega)$ \cite{Jia2018}. In this way, we can coherently map the atomic state $\cos(\theta/2)\ket{G} - \sin(\theta/2)\ket{R}$ onto the free-propagating output photonic state $\ket{\psi_\theta}=\cos(\theta/2)\ket{0}+\sin(\theta/2)\ket{1}$ in a well-controlled mode.

\begin{figure*}[t]
\centering
\includegraphics[width=170mm]{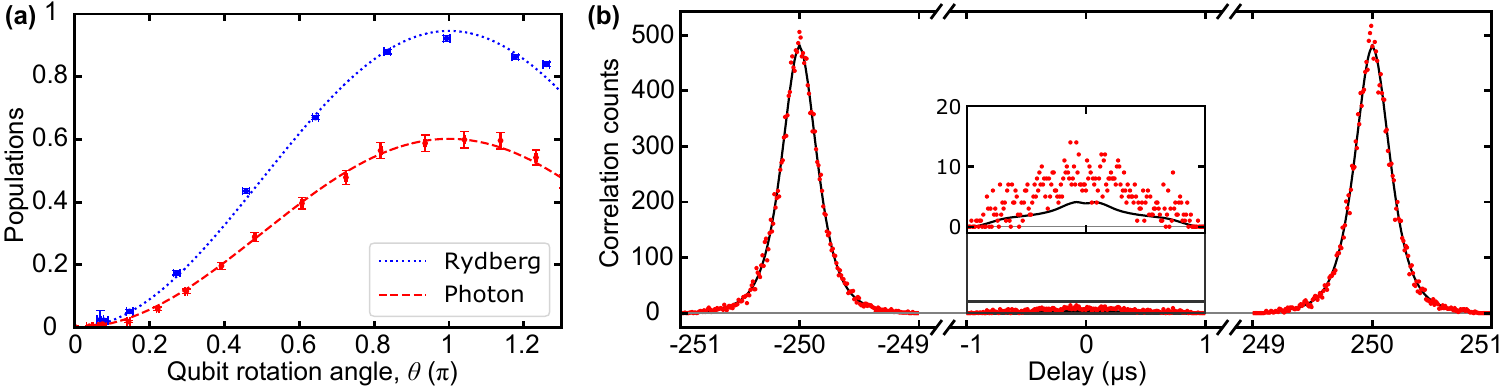}
\caption{ \textbf{High-efficiency single photon generation.} (a) Measured population of states containing at least one Rydberg excitation (blue, details in Sup. Inf.~\ref{app:RydPop}) and measured number of photons in the output pulse (red) while changing the qubit rotation angle $\theta$ via the amplitude of the red write beam. Lines are $\sin^2 \theta$ functions scaled with fitted maximal values of $0.94$ for the Rydberg probability and $0.60$ for the released photon number. (b) Photon number autocorrelation function of the released photonic state with $\theta=\pi$. The inset zooms in around zero delay. Line shows the expected correlations of a perfect single-photon source with the measured experimental background noise.}
\label{fig:EffPhot}
\end{figure*}

We measure the emitted photon flux $I(t)$ with a single photon detector (SPD). Figure~\ref{fig:setup}(d) presents the recorded SPD rate, together with the temporal intensity profiles of the driving laser pulses. Following Ref.~\cite{Stanojevic2011}, the slow variation of the blue read pulse and the long lifetime of the Rydberg state allow us to approximate $I(t)$ as
\begin{equation}
\label{EqApprPhotIntens}
I(t) =  2\kappa \, \eta_C \, \eta_{\mathrm{rem}} \cos^2 b(t)
\exp\left(-2\kappa\int_{t_0}^t \cos^2 b(t') \, \mathrm{d}t' \right),
\end{equation}
where $\eta_C=2C/(2C+1)=93\%$ is our intrinsic Rydberg-to-cavity-photon mapping efficiency limited by the finite cooperativity $C=g^2/(2\kappa\gamma)$ \cite{Gorshkov2007}, $\eta_{\mathrm{rem}}$ is the remaining efficiency term discussed below, $b=\arctan(2g/(\Omega\sqrt{\eta_C}))$, and $t_0$ is the beginning of the read step. For a very large cooperativity, this expression reduces to the leakage rate $2\kappa_D = 2\kappa\cos^2\beta$ of the dark polariton \ket{D} via its photonic component, multiplied by the population $\exp(-2\int_{t_0}^{t}{\kappa_D(t')\mathrm{d}t'})$ remaining in \ket{D} at the time $t$. 
Using the measured intensity of the blue read beam to infer the normalized Rabi frequency profile $\Omega(t)/\Omega_{\mathrm{max}}$ and the values of $g$, $\kappa$ and $\Omega_{\mathrm{max}}=2\pi\times \unit{11.5}{\mega\hertz}$ determined from the transmission spectra in Fig.~\ref{fig:setup}(c), the result of Eq.\ref{EqApprPhotIntens} overlaps with that of a full numerical simulation and with the experimental data, with a slight deviation at short time arising from an artifact in the measurement of $\Omega(t)$.

\subsection*{Efficiency}

Besides $\eta_C$, the overall generation efficiency $\eta=\eta_C \, \eta_{\mathrm{rem}}=\eta_\mathrm{exc} \, \eta_\mathrm{s} \,  \eta_C \, \eta_\mathrm{cav}$ of our device also involves the maximal excitation efficiency $\eta_\mathrm{exc}$ of the state $\ket{R}$, the factor $\eta_\mathrm{s}=e^{-(t_{\mathrm{s}}/\tau_{\mathrm{s}})^2} = 95\%$ describing the thermal dephasing of the Rydberg spin-wave at the time $t_{\mathrm{s}}$ of its retrieval, with a measured characteristic storage time $\tau_{\mathrm{s}} = \unit{2.0}{\micro\second}$ (see Sup. Inf.~\ref{app:Storage}), and the probability $\eta_{\mathrm{cav}} = 90\% $ for an intra-cavity photon to escape through the coupling mirror. As for $\eta_\mathrm{exc}$, it is mostly limited by the finite efficiency of the Rydberg blockade, allowing for a small probability of double Rydberg excitations. Then, the distance-dependent interactions between Rydberg atoms rapidly destroy the phase-matching condition required for an efficient retrieval. This coupling with the quasi-continuum of asymmetric Dicke states \cite{Lukin2001} results in a dephasing rate and in a Lamb shift that can be determined analytically \cite{Magro2023}. Knowing the temporal profiles of our write pulses, they faithfully predict the independently measured maximal probability of $94\%$ to create at least one Rydberg excitation, shown in Fig.~\ref{fig:EffPhot}(a), and they give $\eta_\mathrm{exc}=80\%$. Thus, the calculated value $\eta=63\%$ matches the value of $60 \pm 3\%$ measured with the photon counter and depicted in Fig.~\ref{fig:EffPhot}(a). This efficiency is determined at the output of the superatom-cavity platform where the photonic state is readily available for applications. 

As for the singleness of the emitted photons, the intensity autocorrelation measurement in Fig.~\ref{fig:EffPhot}(b), done with two SPDs in a Hanbury-Brown and Twiss configuration while preparing $\theta=\pi$, shows a very strong antibunching for events within the same pulse compared to consecutive ones. Accounting for the background noise (black curve) and averaging over same-pulse coincidences leads to the autocorrelation $g^{(2)}(0)=0.027$, corresponding to a probability $p(n=2) = 0.005$ for a pulse to contain two photons.

\subsection*{Homodyne tomography}

\begin{figure*}[t]
\centering
\includegraphics[width=170mm]{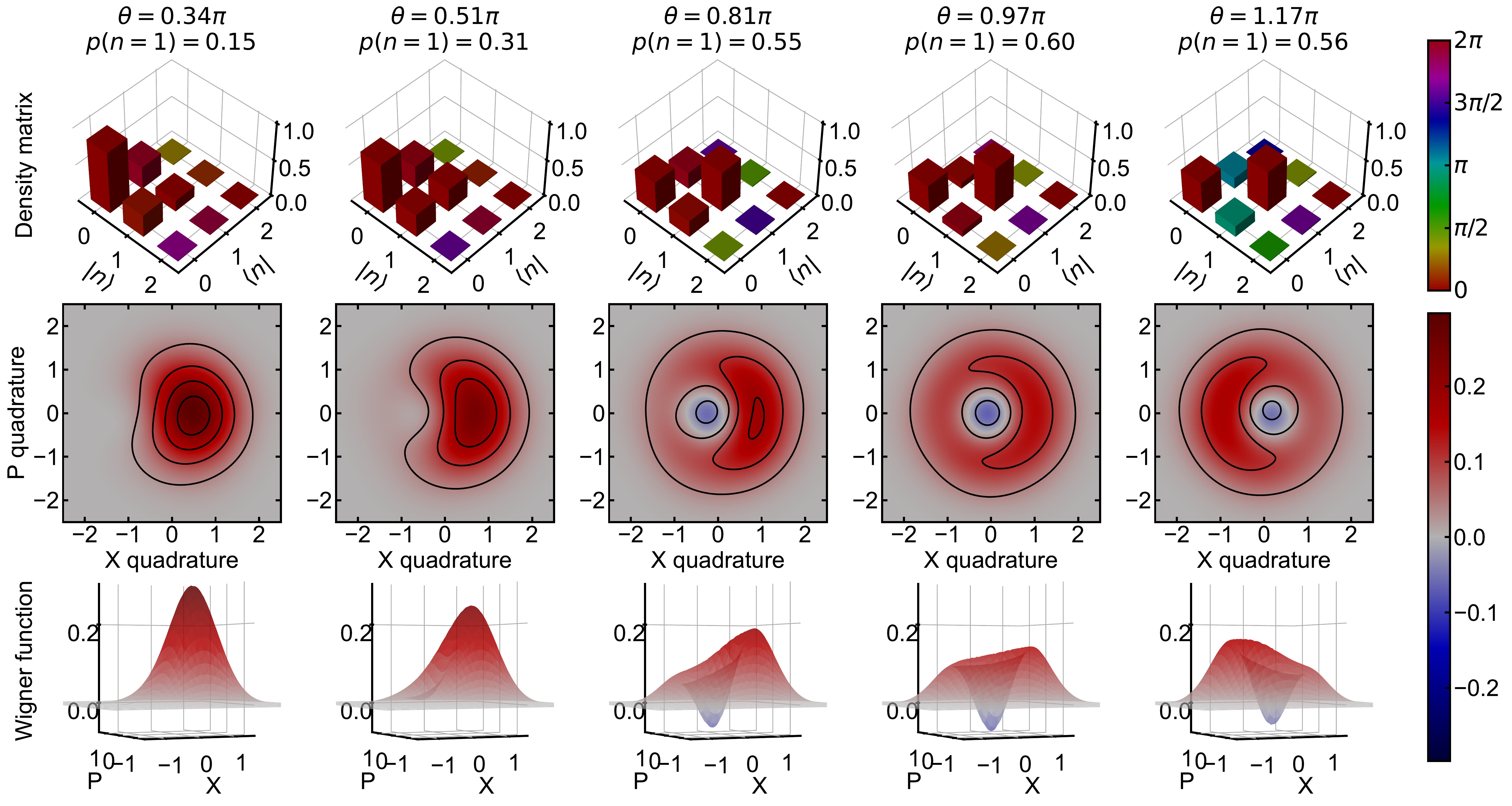}
\caption{ \textbf{Homodyne tomography of photon-number encoded photonic qubits $\ket{\psi_\theta}$,} sorted column-wise by increasing qubit rotation angle $\theta$.  The first row shows the density matrices : for each element, the absolute value is given by the height of the bar, and the argument by its color. Second and third rows show two- and three-dimensional color-mapped representations of Wigner functions over quadrature phase space ($X=\sqrt{2}\mathrm{Re}(\hat{a})$, $P=\sqrt{2}\mathrm{Im}(\hat{a})$).}
\label{fig:HDML}
\end{figure*}

The complete single-mode-selective characterization of the generated photonic states is performed by homodyne tomography, using a shot-noise-limited balanced detector. For each state, we measure the quadrature probability distributions $\mathcal{P}(\hat{X}_\phi=(\hat{a}\mathrm{e}^{i\phi}+\hat{a}^\dag\mathrm{e}^{-i\phi})/\sqrt{2})$ for six phases $\phi_k=k\pi/6$, $k\in[0..5]$, acquiring 32 800 data points per distribution. The mode function $u(t)$, relating the annihilation operator $\hat{a}=\int dt u(t)\hat{a}_\mathrm{out}(t)$ to that of the output field $\hat{a}_{out}(t)$, is taken as the square root of the intensity profile in Fig.\ref{fig:setup}(d). From the distributions $\mathcal{P}(\hat{X}_\phi)$, depicted in Sup. Inf.~\ref{app:quad}, the density matrices $\rho$ and the Wigner functions of the generated states are reconstructed using a maximal likelihood algorithm \cite{Lvovsky2004} including a correction for the detection efficiency.

Figure~\ref{fig:HDML} presents the tomographic reconstructions of five states $\ket{\psi_\theta}$ with different values of the qubit rotation angle $\theta$. When $\theta\approx \pi$, the state resembles a single photon, with an almost-zero ($\rho_{22}=p(n=2)=0.01$) two-photon component of the density matrix. Due to its mostly single-photon character, with $\rho_{11}=p(n=1)=0.60$, the state's Wigner function displays negative values. Within experimental uncertainties, the photon emission efficiency in the mode $\hat{a}$ thus has the same value as the efficiency $\eta$ obtained without temporal mode selection, which highlights the nearly single-mode character of our source and the appropriateness of the chosen mode profile $u(t)$. This is further confirmed by a numerical simulation using the input-output formalism developed in \cite{Kiilerich2019,Kiilerich2020}, which predicts $\rho_{11}=0.59$.

\begin{figure}[t]
\centering
\includegraphics[width=80mm]{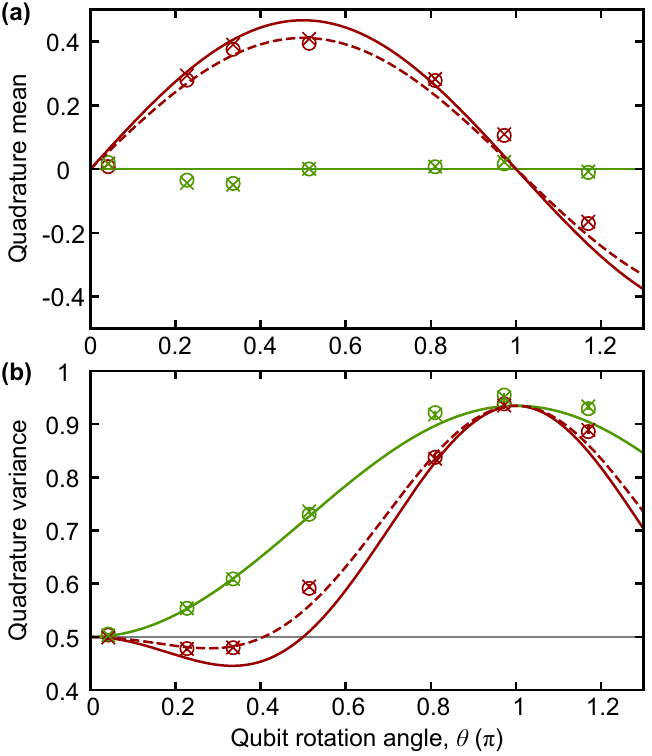}
\caption{ \textbf{Statistics of field quadratures.} Mean value (a) and variance (b) of $X$ (red) and $P$ (green) quadratures of the detected photonic field as a function of the qubit rotation angle $\theta$. Cross data points are extracted directly from quadrature measurements with the corresponding local oscillator phase $\phi$ ($0$ for X, $\pi/2$ for $P$), while circle data points are calculated from the most-likely photonic states inferred from the full tomography data. Full and dashed lines are the expected values without and with a best-fitted pure-dephasing term, respectively. In (b), the gray line marks the variance of the vacuum state. }
\label{fig:HDmeanvar}
\end{figure}

For values of $\theta \neq \pi$, coherences ($\rho_{01}$,$\rho_{10}$) appearing in the density matrices translate into phase-dependent Wigner functions \cite{Resch2002,Lvovsky2002}. While scanning $\theta$ across $\pi$, these coherences flip sign, together with the measured mean value of the $X=X_{\phi=0}$ quadrature (Fig.\ref{fig:HDmeanvar}(a)). As the weight of the single-photon component decreases away from $\theta = \pi$, the negativity of the Wigner function decreases as well. 

At low qubit rotation angles, the state is squeezed along the $X$ quadrature \cite{Carmichael1985}, reaching a $4.4\%$ reduction of the noise variance below the vacuum level without correcting for detection efficiency (Fig.\ref{fig:HDmeanvar}(b)). This value, similar to that obtained with a solid-state artificial atom \cite{Schulte2015}, should ideally reach $25\%$ for $\theta_\mathrm{min}=\pi/3$. Pure losses degrade the squeezing without changing $\theta_\mathrm{min}$, while pure dephasing, stemming from laser frequency noise and fluctuating Stark shifts of the Rydberg resonance, also pushes $\theta_\mathrm{min}$ to lower values. Indeed, pure dephasing has a larger impact on qubit states with important coherences, culminating at $\theta = \pi/2$. Measurements of quadrature means and variances shown in Fig.\ref{fig:HDmeanvar} allow us to determine the pure dephasing rate of $2 \pi \times \unit{0.04}{\mega\hertz}$ by fitting the model of Sup. Inf.~\ref{app:EqQuad}. Such measurements generally provide an elegant alternative to Ramsey spectroscopy for measuring dephasing rates in artificial atoms.

\section*{Discussion}

Our protocol using an intra-cavity Rydberg superatom enables a deterministic generation of single photons, featuring the intrinsic temporal control and  frequency stability of atomic systems. This protocol was developed by combining the advantages of two existing approaches, both relying on strong coupling between a single atomic excitation and a single light mode. The first one consists in using a single trapped atom in a Fabry-P{\'e}rot cavity \cite{McKeever2004,Keller2004}. This method is conceptually simpler but, after two decades of development, such systems struggle to approach unit efficiency due to technical limitations, stemming in particular from residual absorption losses in the required high-finesse cavities \cite{Morin2019,Schupp2021}. A more recent approach uses a Rydberg superatom to obtain a single excitation via the blockade mechanism, subsequently emitted as a single photon with a spatial selectivity provided by the phase-matching of the collective atomic state \cite{Ornelas-Huerta2020}. The efficiency is there physically limited by the achievable optical density of the superatom, which sets the coupling to the output mode. Indeed, reaching a high optical density implies using a high atomic density, leading to a detrimental increase of collisions between the Rydberg electron with ground-state atoms located within its wave function \cite{Schlagmueller2016}. Our approach leverages the combination of a medium-finesse ($F=600$) cavity with a medium-density ($5 \cdot 10^{11}$cm$^{-3}$) superatom to overcome the limits of previous systems. 

Thanks to the obtained state-of-the-art efficiency \cite{Morin2019,Ornelas-Huerta2020,Tomm2021}, we achieved the first fully deterministic preparation of photonic states with negative Wigner functions, reaching a long-sought goal in the quantum optics community \cite{Walschaers2021}. Moreover, the atomic frequency stability combined with a precise control of the optical phases allows us to deterministically produce and measure coherent photon-number-encoded photonic qubits, confirming the potential offered by the emerging field of intra-cavity Rydberg superatoms~\cite{Vaneecloo2022, Jia2018, Yang2022, Stolz2022, Yang2022b}.

Experimental systems such as ours are very recent but they already reached several major milestones in quantum optics \cite{Stolz2022, Yang2022b}, while holding considerable room for improvement.   
Indeed, redesigning the parameters of our cavity specifically for photonic state generation at the expense of some versatility should dramatically increase the free-space mapping efficiency $\eta_C \, \eta_\mathrm{cav}$, thanks to a better photon extraction efficiency and a higher cooperativity. Higher excitation efficiencies $\eta_\mathrm{exc}$ and lower dephasing rates between the states \ket{G} and \ket{R} could be reached via more involved laser pulse sequences \cite{Beterov2013,Omran2019}. There is every reason to believe that this approach, combined with state-of-the-art single-photon counters, could exceed the source-detector efficiency threshold of $2/3$, required for efficient linear optical quantum computation \cite{Varnava2008}.
Generally, the abilities of intra-cavity Rydberg superatom systems remain largely unexplored. They hold promises as sources of sophisticated non-classical light states or as photonic quantum simulators, enabled by coupling several Rydberg superatoms to the same cavity mode or by coupling several cavity modes via Rydberg interactions \cite{Clark2020}. The flexibility and the controllability of our platform make it an excellent testbed to explore such schemes and their applications for optical quantum engineering.

\section*{Methods}

\subsubsection*{Superatom preparation}

The ensemble of cold atoms is produced in $\unit{0.1}{\second}$ by the sequential use of two-dimensional (2D) magneto-optical trapping (MOT), 3D MOT, optical molasses cooling, trapping and transporting in a one dimensional optical lattice, degenerate Raman sideband cooling and trapping in a crossed dipole trap (CDT), described in details in Ref.~\cite{Vaneecloo2022}. Here, we follow it by an adiabatic ramp-up of the CDT beams power by a factor $2.5$ to reduce the cloud's radius. In the last initialization stage, the atoms are pumped to the state $G: 5S_{1/2}, F\!\!=\!\!1, m_F\!\!=\!\!1$ to obtain a collective coupling strength $g = 2\pi \times \unit{10}{\mega\hertz}$ between the superatom and the cavity field. Then, for $41$ times (or $101$ times for data in Fig.~\ref{fig:EffPhot}(b)), we repeat write and read operations, followed by an electric field pulse applied on in-vacuum electrodes to remove residual Rydberg atoms, and by a repumping of reservoir atoms from uncoupled state $5S_{1/2}, F\!\!=\!\!2$ to $G$ in order to prevent a decrease of $g$ from atom losses. During the write and read operation, we momentarily switch off the CDT to avoid differential light-shifts detrimental for superatom coherence~\cite{Schmidt2020}.

\subsubsection*{Detections efficiencies}

We infer the measured value of the generation efficiency $\eta$ from the measured SPD detection probability per generation cycle and the measured detection efficiency of $40.5 \pm 1.6 \%$. The main contributions to the latter arise from the diffraction efficiency of detection AOM ($75\%$, see Fig.\ref{fig:setup}), the coupling in a singlemode fiber ($89\%$) and the detector quantum efficiency ($68\%$), with smaller contributions from optical elements, in particular from interference filters protecting the SPD from stray light. The tomography results from the maximum likelihood algorithm account for the overall $72.2 \pm 2.3 \%$ efficiency of the homodyne detection. It includes optical losses in the path towards the detector through the AOM and an optical isolator ($91\%$), the mode matching with the local oscillator ($92.6\%$), the photodiodes quantum efficiencies ($91\%$) and the electronic noise ($98\%$). 
The quite large relative uncertainties of the detection efficiencies stem mostly from conservative estimates on the uncertainties of the quantum efficiencies of the detectors.

\subsubsection*{Coherent 4-photon process}

Our generation of a phase-coherent superposition of \ket{0} and \ket{1} photonic states relies on a precise control of the optical phases of the write and read beams, relative to phase reference provided by the local oscillator (LO). The blue beam frequencies, on one side, and the red write and LO frequencies, on the other side, both differ by the detuning of $\unit{500}{\mega\hertz}$, imposed by the free spectral range of the confocal amplification cavity increasing the intensities of the blue beams \cite{Vaneecloo2022}. We use a $\unit{500}{\mega\hertz}$ radio frequency, on one side, to demodulate the beat note between the blue write and read beams recorded with a fast photodiode, and, on the other side, to drive a fibered electro-optical modulator (EOM) on the LO path allowing to measure the red write phase with the homodyne detector. Both signals are fed to the respective AOM driving phases via a digital controller operating in a lock-and-hold regime, applied at the end of the superatom initialization step. By applying this technique, we measured standard deviations of $\unit{1.4}{\degree}$ between the blue phases and $\unit{6}{\degree}$ between the red write and the LO phases. This lock ensures the sequence-to-sequence stability of the coherent phase of the photonic state. 
However, the phase of the single-photon field also varies over the duration of its emission. This is due, on one hand, to the time-dependent light shift created by the blue read beam on the $G-E$ transition and, on the other hand, to the time-dependent phase of the blue read beam, resulting from an interplay between the response of the dedicated AOM and the buildup of the field inside the amplification cavity. This dynamic phase is compensated by applying a matching profile to the LO phase via the EOM (see Sup. Inf.~\ref{app:PhotPhase}).

\section*{Acknowledgments}

This work was funded by the ERC Starting Grant 677470 SEAQUEL and the CIFAR Azrieli Global Scholars program. The authors thank P. Travers for technical support, and S. \'Cuk and M. Enault-Dautheribes for their assistance at the early stage of the project.

\clearpage

\newpage

\appendix

{\LARGE Supplementary Information}

\section{Measurement of Rydberg probability}
\label{app:RydPop}

A key parameter characterizing our system is the probability of inserting a Rydberg excitation in the atomic cloud. To measure it, we use the Rydberg blockade effect: the presence of a Rydberg excitation in the cloud destroys the electromagnetically-induced transparency window observed in the resonant transmission of the cavity on Fig.~\ref{fig:setup}(c). Measuring this transmission allowed us to perform a high-fidelity single-shot detection of a Rydberg excitation in Ref.~\cite{Vaneecloo2022}.

\begin{figure}[t]
\centering
\includegraphics[width=70mm]{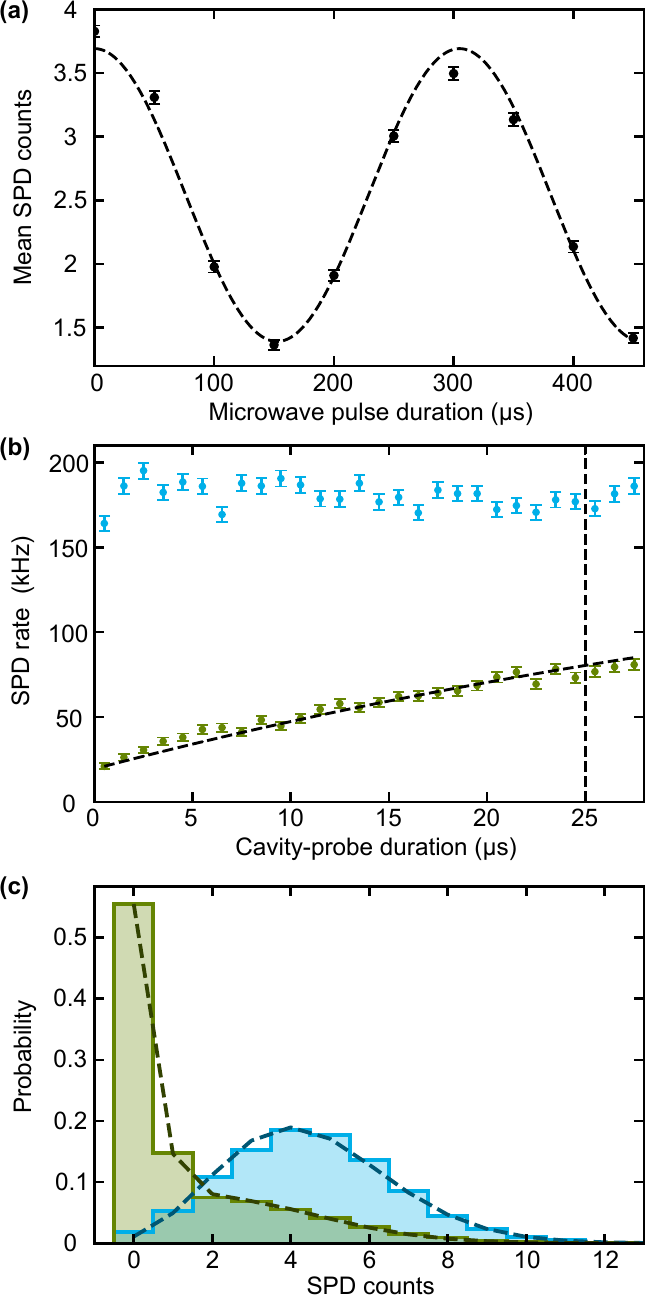}
\caption{ \textbf{Measurement of Rydberg population.} (a) Calibration of microwave $\pi$ pulse to transfer between Rydberg states $109 S_{1/2}$ and $109 P_{3/2}$. While changing the pulse duration, Rabi oscillations are visible on the mean SPD counts, integrated over \unit{25}{\micro\second}, of the following optical detection of Rydberg EIT. The dashed line is a sinusoidal best-fit function. (b) SPD count rate while probing the cavity transmission at the dark polariton resonance, either after a $\pi$ write pulse and a $\pi$ microwave pulse (green) or without write pulse (cyan). The dashed line is a best-fit exponential decay with a lifetime $\tau_R = \unit{53}{\micro\second}$. The vertical dashed line indicates the upper limit of the integration time window. (c) Histograms of the number of SPD counts, within the integration window, without (cyan) and with (green) a $\theta=\pi$ write pulse. The dashed curves represent modeled histograms}
\label{fig:RydPop}
\end{figure}

In order to detect state of the superatom after the write step as presented in Fig.~\ref{fig:RydPop}(a), we use here a modified scheme requiring only a single blue laser \cite{Xu2021}.
 We transfer the Rydberg excitation from the $109S$ to the $109P_{3/2}$ state using a microwave pulse at the resonant frequency of \unit{2.694}{\giga\hertz}. We apply the microwave pulse \unit{500}{\nano\second} after the write pulse, which is fast enough to neglect the decay of the $109 S$ state, which has a \unit{426}{\micro\second} calculated lifetime in a \unit{300}{\kelvin} environment \cite{Sibalic2017}. \unit{6.7}{\micro\second} after the microwave pulse, the detection beams are activated for \unit{30}{\micro\second}. The crossed dipole trap beams are switched back on about \unit{0.7}{\micro\second} earlier to avoid loosing atoms during the detection. The detection uses a \unit{795}{nm} $D1$-resonant beam transmitted through the cavity, as well as the blue read beam. The mean detected photon number during the first \unit{25}{\micro\second}, shown on Figure~\ref{fig:RydPop}(a), reflects the Rabi oscillations between states $109 S$ and $109 P_{3/2}$. Here, the superatom is initially prepared to reach a maximal $109S$ population with a $\theta = \pi$ write pulse.
The \unit{160}{\nano\second} microwave $\pi$ pulse is calibrated accordingly to transfer the Rydberg excitation from $109 S$ to $109 P_{3/2}$. 

When the superatom is in the ground state $\ket{G}$, the EIT enables a high and constant photon flux  $\phi_G$ through the cavity as depicted in Fig.~\ref{fig:RydPop}(b) by the cyan data. When a $\theta=\pi$ write pulse creates a Rydberg excitation inside the cloud, the transmission drops down as illustrated by the green data points. It increases with time due to the decay of the Rydberg excitation back to the ground state, with an effective lifetime $\tau_R = \unit{53}{\micro\second}$.

The related statistical histograms of detected SPD counts during the first \unit{25}{\micro\second}, presented in Fig.~\ref{fig:RydPop}(c), show a clear separation depending on absence or presence of a a Rydberg atom. 
Hence, to estimate the Rydberg presence probability $\zeta_R$ for a given write pulse (Fig.~\ref{fig:EffPhot}(a) in the main text), we use the model described in Ref.~\cite{Vaneecloo2022} to fit the measured histograms. This model takes into account quantum jumps during the integration and yields a probability to observe a photon number $n$ as $\zeta_R \mathcal{P}_R(n)+(1-\zeta_R)\mathcal{P}_G(n)$ with
\begin{eqnarray}
	\mathcal{P}_R(n) &=& \mathcal{P}(n,  t_i \phi_R)e^{-t_i/\tau_R}\\  \nonumber 
	& &+ \int_{0}^{t_i} \mathcal{P}(n, t \phi_R + (t_i-t) \phi_G) \frac{e^{-t/\tau_R}}{\tau_R} dt 
\end{eqnarray}
and $\mathcal{P}_G(n) = \mathcal{P}(n, t_i \phi_G)=e^{-t_i \phi_G}(t_i \phi_G)^n/n!$ a poissonian law. Knowing the mean count $t_i \phi_G = 4.5$ without a write pulse and the lifetime $\tau_R = \unit{53}{\micro\second}$, the probabilty distribution is fitted with only the residual transmission in presence of Rydberg $\phi_R$ and the Rydberg probability $\zeta_R$ as free parameters. The best fitted values of $\zeta_R$ are reported in Fig.~\ref{fig:EffPhot}(a) with a correction factor of $1/0.987$ accounting for the decay of the $109P_{3/2}$ state, with a \unit{509}{\micro\second} lifetime, over the \unit{6.7}{\micro\second} delay between the microwave pulse and the start of the detection. 

\section{Storage time}
\label{app:Storage}

\begin{figure}[t]
\centering
\includegraphics[width=80mm]{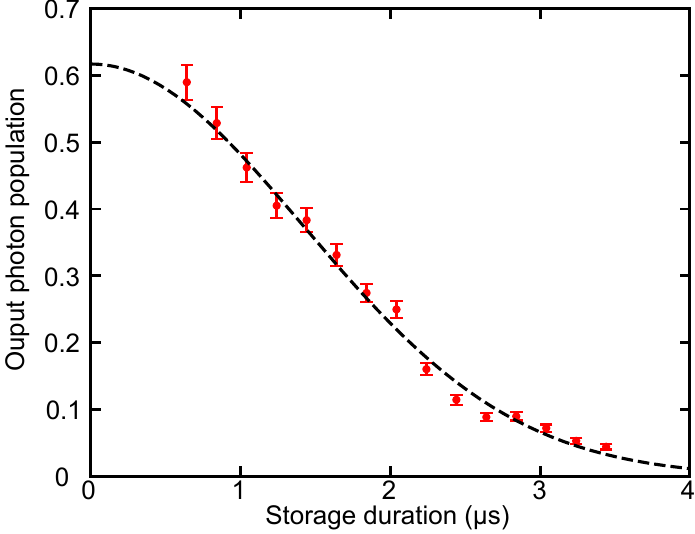}
\caption{\textbf{Storage time.} Number of photon per pulse extracted from the cavity as a function of the storage duration between write and read steps. The dashed line is a best-fit Gaussian decay, yielding the storage time $\tau_s = \unit{2.0}{\micro\second}$.}
\label{fig:StorageT}
\end{figure}

The preparation of a single photon requires two steps in our scheme: the first one stores a Rydberg excitation $\ket{R}$, then the second one maps this excitation on a photonic state. Between these steps, the retrieval efficiency decays with a characteristic time dubbed the storage time $\tau_s$.
In this section, we focus on the physical origin of this decay and on the measurement of $\tau_s$. The storage time measurement consists in sending a $\pi$ writing pulse on the atomic cloud then waiting a time $t_s$ before reading it out. We define $t_s$ as the difference between the mean times of the red write pulse and of the created photon, using their temporal profiles as weighting functions. This measurement gives access to the nature of the decay processes acting on the state $\ket{R}$, which corresponds to a single Rydberg excitation delocalized over the whole ensemble. 
The contribution of an atom at the position $\vec{r}_n$ to this superposition carries a phase factor $\exp(i \vec{k}_{GR}\cdot \vec{r}_n)$ which involves the sum of the wavevectors of the red and blue excitation beams  $\vec{k}_{GR}$. 
The non-zero temperature $T$ inside the cloud induces a motional dephasing which leads to a Gaussian decay  $e^{-t_s^2/\tau_{T}^2}$  with a characteristic time $\tau_{T} = \sqrt{m/k_{\mathrm{B}} T} / \left\|\vec{k}_{GR}\right\|$, where $m$ is the mass of the atom, $k_{\mathrm{B}}$ is the Boltzmann constant. With a calculated characteristic time of \unit{3}{\micro\second}, the measured cloud's temperature of $T=\unit{5}{\micro\kelvin}$ is thus the main contribution to the Gaussian storage time $\tau_s =  \unit{2}{\micro\second}$ obtained by fitting the experimental data in Fig.~\ref{fig:StorageT}. 

\section{Dynamic photon phase}
\label{app:PhotPhase}

\begin{figure}[t]
\centering
\includegraphics[width=80mm]{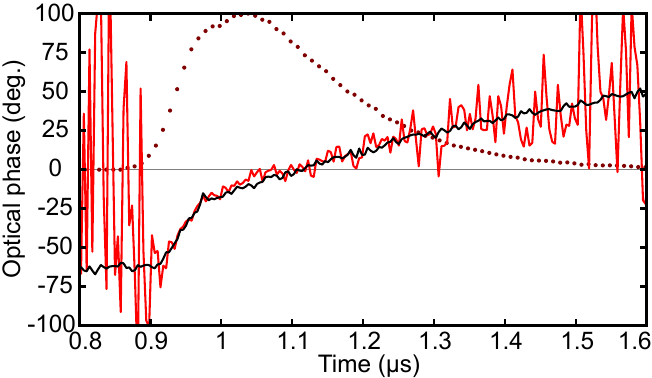}
\caption{\textbf{Measured photon phase.} Phase of the output photonic mode (red) measured with the bare homodyne detection, and corresponding measured compensation on the detection phase (black) via a temporal modulation of the local oscillator phase. The dark-red data points are the rescaled measured SPD rate to indicate the intensity profile of the output photonic mode. The time reference is identical to Fig.~\ref{fig:setup}(d).}
\label{fig:LSphase}
\end{figure}

\begin{figure*}[t]
\centering
\includegraphics[width=170mm]{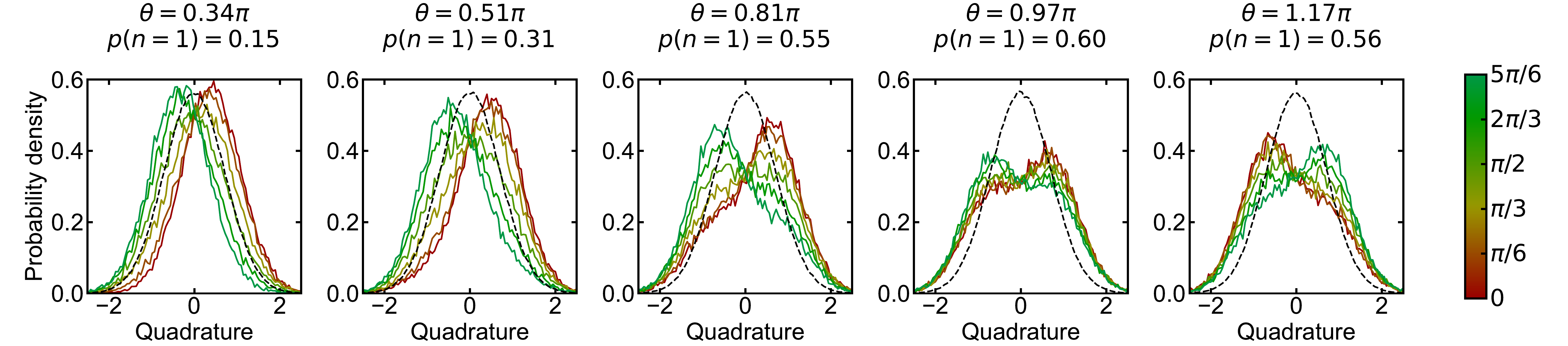}
\caption{\textbf{Tomography bare results.} Quadratures histograms obtained from homodyne detection measurements, with the colors corresponding to the LO phase set between $0$ and $5\pi/6$ with $\pi/6$ steps. Black dashed histograms are the Gaussian vacuum noise references with $1/\sqrt{2}$ standard deviations. From this data, a maximum-likelihood algorithm reconstructs the states presented in Fig.~\ref{fig:HDML}. Each column, corresponding to an increasing qubit rotation angle $\theta$, indicates the population of the single-photon state $p(n=1)$. }
\label{fig:HDhistos}
\end{figure*}

The homodyne detection is based on frequency matching between the local oscillator (LO) and the photonic state of interest. Yet, as shown in Fig.~\ref{fig:LSphase}, the optical phase of our photonic qubit drifts over the duration of its emission. For this measurement, we use a $\theta=\pi/2$ qubit rotation angle, which gives a good trade-off between a strong photonic field amplitude and a well-defined optical phase, and we keep the LO at a constant phase. The optical phase is then retrieved from the averages of the X and P quadratures. The measured total phase drift is around \unit{125}{\deg} which, in frequency domain, translates in a \unit{350}{\kilo\hertz} detuning with the LO. The linear component of this dynamic phase results from a slight frequency mismatch in the calibration of the two-photon resonance. The non linear drift, visible mostly at the beginning of the photon emission, can be explained by the combination of two effects. First, the bright read beam shifts the two-photon $G-R$ resonance, due to its far off-resonant coupling with the $G-E$ transition. However, according to the measured light shift of the dark polariton resonance, this process alone cannot explain the magnitude of the dynamic nonlinear phase shift of the output photon. We attribute the remaining dynamics to transient spatial inhomogeneities of the read blue beam. Indeed, when the blue read AOM switches on, the shape and the phase of the diffracted beam vary. Inside the confocal blue amplification cavity, the mode and the phase of this beam transiently vary as well, the variations of the phase getting imprinted on that of the output photon. 

To optimally detect the output photonic state, we shape the LO phase during the detection to compensate the phase variations of the emitted photonic field. To control the phase of the LO, we use the same fibered EOM which creates a sideband shifted by \unit{500}{\mega\hertz} for the phase locking step (see Methods). The EOM voltage is dynamically controlled with a \unit{250}{\mega\hertz} sample rate by an arbitrary waveform generator spanning the $V_\pi=\unit{4}{\volt}$ voltage of the EOM with a 14-bit resolution. The optical phase of the local oscillator is detected using the write red beam which is resonant with the sideband. The black curve in Fig.~\ref{fig:LSphase} is sign-reversed to assess its agreement with the phase drift of the output photonic state. The efficiency of this compensation is confirmed by the agreement between the photon emission efficiencies measured with the SPD and with the homodyne detection: without this phase matching, the mode selected by the homodyne detection is sub-optimal and the respective efficiency becomes lower than the one measured with the SPD.

\section{Homodyne tomography process}
\label{app:quad}

Raw data from the homodyne acquisition are first filtered by the square root of the temporal mode of the output photon (Fig.~\ref{fig:setup}(d)) then gathered in histograms displayed in Fig.~\ref{fig:HDhistos}. For each angle $\theta$ of the qubit superposition, a maximum likelihood algorithm  is then implemented, taking into account detection losses and leading to Fig.~\ref{fig:HDML}. This alogrithm uses 500 optimization cycles and considers a Hilbert space limited to 5 photons.

\section{Dephasing effect on output quadratures}
\label{app:EqQuad}

Between the write and the read steps, the superatom, initially prepared in a superposition of ground and Rydberg states $\cos(\theta/2)\ket{G}-\sin(\theta/2)\ket{R}$, evolves according to the master equation
\begin{eqnarray}
	\frac{\dd}{\dd t}\orho &=& \mathcal{L}[\sqrt{2\gamma_1}\os](\orho) +\mathcal{L}[\sqrt{2\gamma_2}\osd\os](\orho)\label{EqMaster}\\
	\mathcal{L}[\oA](\orho) &=& \oA\orho\oAd-\frac{1}{2}\orho\oAd\oA-\frac{1}{2}\oAd\oA\orho \nonumber
\end{eqnarray}
with $\os=\ket{G}\bra{R}$, $\gamma_1$ being the homogeneous decay and $\gamma_2$ a pure dephasing term. The readout process maps atomic states $\ket{G}$ and $\ket{R}$ onto Fock states $\ket{0}$ and $\ket{1}$ respectively. It is accompanied by linear losses due to a finite cooperativity and to a finite extraction efficiency. These losses can be grouped together and modeled by a beamsplitter with a transmission $\zeta$ mixing the retrieved field $\oa$ with a vacuum mode $\ob$: $\oa  \rightarrow \sqrt{\zeta}\oa+\sqrt{1-\zeta}\ob$. By integrating the equation \ref{EqMaster} then adding linear losses one can find the means and the variances of the quadratures $\oX$ and $\oP$ of the output field,
\begin{eqnarray}
	\langle\oX\rangle &=& \sqrt{2 \eta}\ee^{-\gamma_2 t_s} \sin\left(\frac{\theta}{2}\right) \cos\left(\frac{\theta}{2}\right)\nonumber\\
	\langle\oP\rangle &=& 0\nonumber\\
	\langle\Delta\oX^2\rangle &=& \eta\sin^2\left(\frac{\theta}{2}\right)\left[1-2\ee^{-2\gamma_2 t_s}\cos^2\left(\frac{\theta}{2}\right)\right]+\frac{1}{2}\nonumber\\
	\langle\Delta\oP^2\rangle &=&  \eta\sin^2\left(\frac{\theta}{2}\right)+\frac{1}{2}\nonumber
\end{eqnarray}
with $\eta=\zeta \ee^{-2\gamma_1 t_s}$ where $t_s = \unit{0.48}{\micro\second}$ is the storage duration. In Fig.~\ref{fig:HDmeanvar} dashed lines are the expected values with a pure-dephasing term, $\gamma_2 = 2 \pi \times \unit{0.04}{\mega\hertz}$. This parameter is the average of the values obtained from the best fits on the quadrature means and variances. 

\end{document}